\documentclass[12pt]{article}

\usepackage[dvips]{graphicx}
\usepackage{cite}
\usepackage{amsmath}
\usepackage{amssymb}
\usepackage{subfigure}
\usepackage{wasysym}
\usepackage{bbold}

\usepackage[dvips]{graphicx}
\usepackage{cite}
\DeclareGraphicsExtensions{.eps}

\newcommand{\ri}{{ \rm i }}

\newcommand{\rd}{{ \rm d }}
\newcommand{\be}{\begin{equation}}
\newcommand{\ee}{\end{equation}}

\newcommand{\Ai}{{\rm Ai}}

\newcommand{\sech}{{\rm sech}}

\begin{document}


\title{Uniform semiclassical approximations of the nonlinear Schr\"odinger
equation by a Painlev\'e mapping}
\author{D. Witthaut and H. J. Korsch\\
FB Physik, Technische Universit\"at Kaiserslautern \\
D--67653 Kaiserslautern, Germany}
\date{\today}

\maketitle

\begin{abstract}
A useful semiclassical method to calculate eigenfunctions of the
Schr\"odinger equation is the mapping to a well-known ordinary
differential equation, as for example Airy's equation.
In this paper we generalize the mapping procedure to the nonlinear
Schr\"odinger equation or Gross-Pitaevskii equation describing
the macroscopic wave function of a Bose-Einstein condensate.
The nonlinear Schr\"odinger equation is mapped to the second
Painlev\'e equation ($P_{II}$), which is one of the best-known
differential equations with a cubic nonlinearity.
A quantization condition is derived from the connection formulae
of these functions. Comparison with numerically exact results for
a harmonic trap demonstrates the benefit of the mapping method.
Finally we discuss the influence of a shallow periodic potential on
bright soliton solutions by a mapping to a constant potential.
\end{abstract}


\section{Introduction}

In the case of low temperatures, the dynamics of a Bose-Einstein condensate (BEC)
can be described in a mean--field approach by the nonlinear Schr\"odinger
equation (NLSE) or Gross--Pitaevskii equation (see, e.g., \cite{Pita03})
\be
  \left( - \frac{\hbar^2}{2M} \frac{\partial^2}{\partial x^2} + V(x) + g |\psi(x,t)|^2 \right)
  \psi(x,t) = \ri \hbar \frac{\partial \psi(x)}{\partial t} \, ,
  \label{eqn-NLSE-timedep}
\ee
where $g$ is the nonlinear interaction strength. Stationary nonlinear
eigenstates of the NLSE satisfying $\psi(x,t) = \exp(-\ri \mu t/\hbar) \psi(x)$
fulfill the time-independent NLSE
\be
  \left( - \frac{\hbar^2}{2M} \frac{\rd^2}{\rd x^2} + V(x) + g |\psi(x)|^2 \right)
  \psi(x) = \mu \psi(x) \, .
  \label{eqn-NLSE-general}
\ee
Analytic solutions of the NLSE are available only for some special cases,
among these the free NLSE \cite{Carr00a,Carr00b}, arrangements of delta-potentials \cite{04nls_delta,05dcomb,Seam05} and potentials given by Jacobi elliptic
functions \cite{Bron01a}. Thus there is a great interest in feasible
approximation to the NLSE, among which the most popular one is the
Thomas-Fermi approximation for the ground state in a trapping potential.

Recently two semiclassical methods for the calculation of nonlinear
eigenstates were proposed.
Mention and discuss other semiclassical approaches to the NLSE:
Bohr-Sommerfeld quantization \cite{Kono03} and the divergence-free
WKB method \cite{Hyou02,Hyou04}

In this paper we discuss another semiclassical methods to solve the
time-independent NLSE approximately based on a mapping to Painlev\'e
second equation.
Special attention will be paid to the shift of the chemical potential $\mu$
of bound states due to the nonlinear mean-field energy. This method
provides good results for excited states and it is fairly easy to
understand and to use.

We will focus on the one--dimensional case, which arises, for example,
in confined geometries (see, e.g., \cite{Grei01} and references therein).
Then the effective interaction strength is given by
$g = 2 \hbar^2 a N / M a_\perp^2$, where $a$ is the s-wave scattering length,
$a_\perp$ is the transverse extension of the condensate and $N$ is the number of
atoms \cite{Olsh98}. We assume that the wave function is normalized
as $\| \psi \|^2 = 1$.

For convenience we rescale the NLSE (\ref{eqn-NLSE-general}) such that
$\hbar = M = 1$.This yields the NLSE in the convenient form
\be
   \frac{\rd^2 \psi}{\rd x^2} = - q^2(x) \psi(x) + 2 g |\psi(x)|^2 \psi(x)
   \label{eqn-nlse-standard}
\ee
with $q^2(x) = 2(\mu - V(x))$.

For the most important case of a harmonic trap $V(x) = m\omega^2 x^2/2$
discussed in section \ref{sec-harmosc} this is achieved by the rescaling
the variables as
\be
  x' = x/\ell, \quad
  \psi' = \sqrt{\ell} \psi, \quad
  g' = \ell M g/ \hbar^2 \quad
  \mbox{and} \, \mu' = \mu/(\hbar \omega)
\ee
with the standard length $\ell = \sqrt{\hbar/M \omega}$. The rescaled potential
is $V(x) = x^2/2$ and the chemical potential is given in units of $\hbar \omega$.
To get a feeling for the relevant dimensions consider a BEC of $10^4$
atoms with transverse width $a_\perp = 10 \mu m$.
This yields a scaled nonlinearity of $g = +20$ for a ${}^{87}{\rm Rb}$
BEC in a trap with axial frequency $\omega = 2\pi \times 2 {\rm Hz}$
and $g = -20$ for a ${}^7{\rm Li}$-BEC in a trap with
$\omega = 2\pi \times 100 {\rm Hz}$.

Finally, let us note that the nonlinear Schr\"odinger equation also
describes the propagation of electromagnetic waves in nonlinear media
(see, e.g., \cite{Dodd82}, ch. 8).

\section{The second Painlev\'e transcendent}

One of the most famous ordinary differential equation with a cubic
nonlinearity is Painlev\'e's second equation or shortly the $P_{II}$
equation (see, e.g. \cite{Iwas91,Ablo91}),
\be
  \frac{\rd^2 \phi}{\rd y^2} = 2 \sigma \phi^3 + y \phi \, , \quad \sigma = \pm 1.
  \label{eqn-p2}
\ee
The solutions $\phi_k(x)$, where the index $k$ refers to the asymptotics
at $y \rightarrow + \infty$, are transcendent.
In the linear case, which is found for $\sigma = 0$, the $P_{II}$ equation
reduces to the Airy equation. In fact Airy functions are found in the
asymptotic limit (see below).

In textbooks one mostly finds results for the repulsive case $\sigma = +1$.
However, asymptotic expansions, which will prove itself as quite useful,
are available for both cases \cite{Ablo77,Segu81}.
For $y \rightarrow + \infty$ the Painlev\'e transcendent $\phi_k(x)$ vanishes as
\be
  \phi_k(y) \sim k \Ai(y)
  \label{eqn-p2-airy}
\ee
We can restrict ourselves to $k > 0$, since equation (\ref{eqn-p2}) is
invariant under a sign change of $\phi(y)$.
Connection formulae, which relate the asymptotic form for
$y \rightarrow - \infty$ to the form for  $y \rightarrow + \infty$
are well known \cite{Ablo77,Segu81}.
For $\sigma = +1$ and $k \ge 1$ the Painlev\'e transcendent diverges.
Otherwise the solution is oscillatory for negative $y$ with asymptotics
\be
  \phi_k(y) \sim d |y|^{-1/4} \sin\left( \dfrac{2}{3} |y|^{3/2} -
  \dfrac{3}{4} \sigma d^2 \ln |y| - \theta \right) + \mathcal{O} (|y|^{-7/4}).
  \label{eqn-p2-asymp2}
\ee
The constants depend on the parameter $k$ as
\begin{eqnarray}
  d^2(k) &=& - \frac{\sigma}{\pi} \ln(1- \sigma k^2)
  \label{eqn-p2-connect-const1} \\
  \theta(k) &=& \frac{3}{2} \sigma d^2(k) \ln 2 +
  \sigma \arg \left[\Gamma\left(1-\dfrac{1}{2} \ri d^2(k)\right)\right] - \dfrac{\pi}{4}.
  \label{eqn-p2-connect-const2}
\end{eqnarray}

\begin{figure}[t]
\centering
\includegraphics[width=6cm,  angle=0]{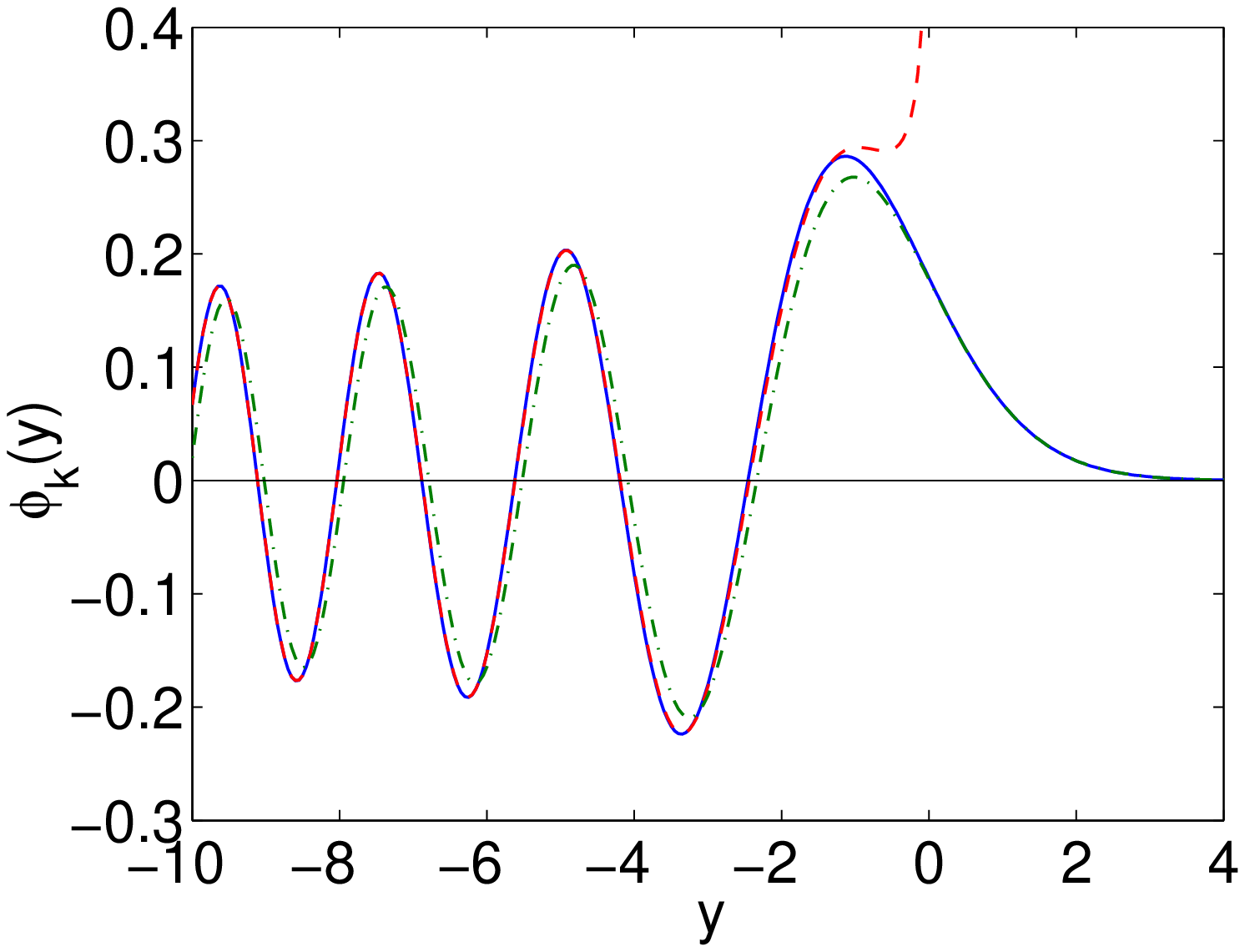}
\hspace{5mm}
\includegraphics[width=6cm,  angle=0]{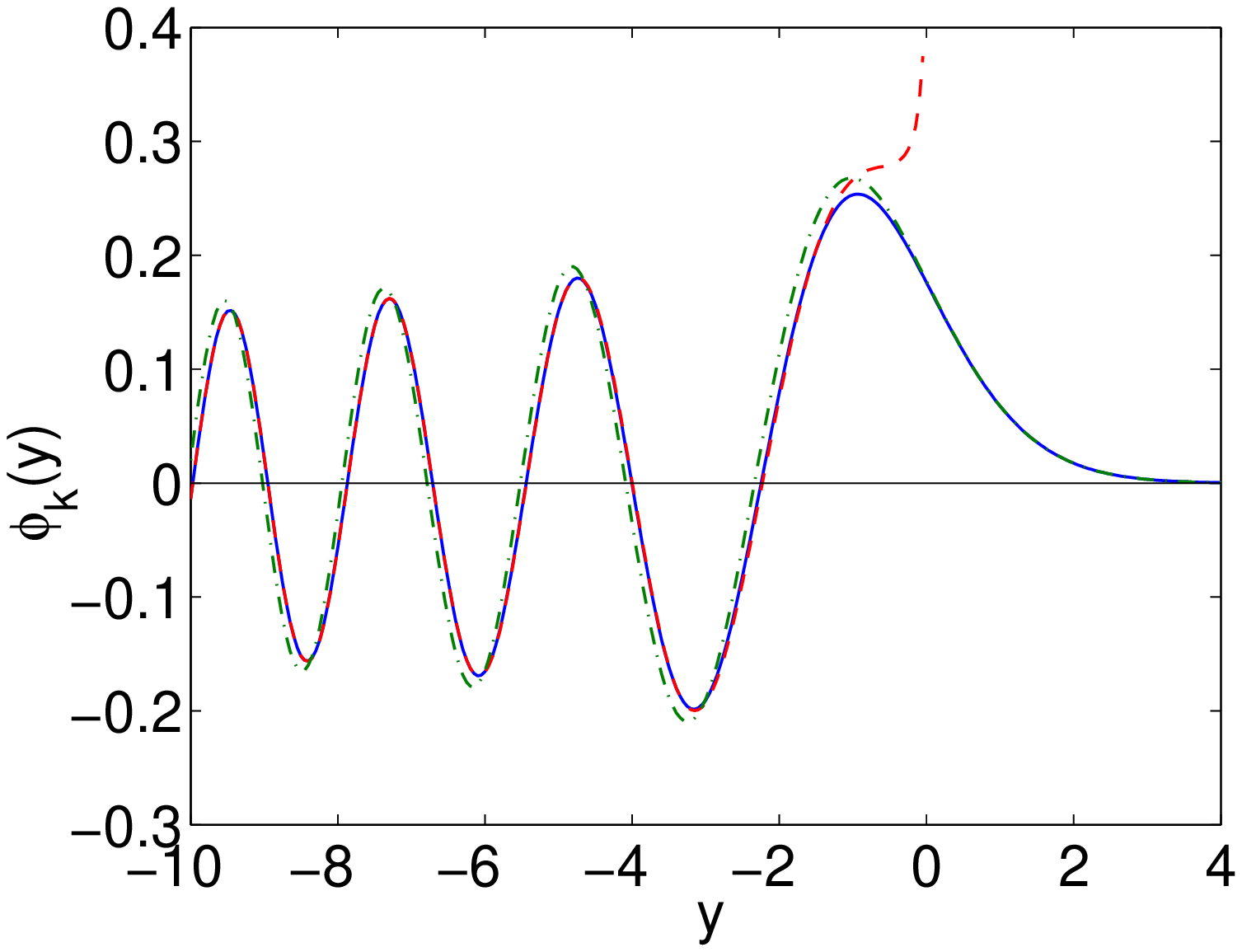}
\caption{\label{fig-p2_k=0.5}
The Painlev\'e transcendent $\phi_k(y)$ (solid blue lines) for $k = 0.5$
and $\sigma = +1$ (left) and $\sigma = -1$ (right) in comparison with the
asymptotic forms (\ref{eqn-p2-asymp2}) (red dashed line) and
$k \Ai(y)$ (green dash-dotted line).}
\end{figure}

The form of the second Painle\'e transcendent is illustrated in figure
\ref{fig-p2_k=0.5}.
We plotted the Painlev\'e transcendent $\phi_k(y)$ in comparison with the
asymptotic expansions (\ref{eqn-p2-asymp2}) for $y<0$ and (\ref{eqn-p2-airy})
for $y>0$ for $k = 0.5$ and $\sigma = \pm 1$. One observes that the asymptotic
expansions are quite accurate already for small values of $|y|$.

Furthermore, note that the $P_{II}$ equation with $\sigma = +1$ can be written
as a Hamiltonian system \cite{Ablo91}
\be
  \frac{\rd \phi}{\rd x} = \frac{\partial {\cal H}}{\partial p} \, , \qquad
  \frac{\rd p}{\rd x} = - \frac{\partial {\cal H}}{\partial \phi}
\ee
with the Hamiltonian function
\be
  {\cal H} = \frac{p^2}{2} - \left( \phi^2 + \frac{x}{2} \right) p - \frac{\phi}{2} .
\ee

\section{The wedge potential}
\label{sec-wedge}

As a first illustrative example for the application of the $P_{II}$
equation  we consider the real-valued nonlinear eigenstates  in a
wedge potential
\be
  V(x) = F |x|.
\ee
This potential might appear a bit artificial, but it provides
a natural and  easily understandable example for the
nonlinear quantisation using the $P_{II}$ transcendent.
Furthermore, the quantum states of cold neutrons in the earth's
gravity potential above a hard wall corresponding to a half-wedge
were measured only recently \cite{Nesv02}.

We consider only real states with a defined parity
$\psi(x) = (-1)^n \psi(-x)$, such that we can restrict our analysis
to the positive real line, $x >0$ and replace $|\psi(x)|^2 \psi(x)$
by $\psi(x)^3$
By the means of a scaling $y = (2F)^{1/3} (x - \mu / F )$ and
$\psi = |g|^{-1/2} (2F)^{1/3} \phi$, the NLSE
with the wedge potential is transformed to the standard form
\be
  \frac{\rd^2 \phi}{\rd y^2} = 2 \sigma \phi^3 + y \phi
  \label{eqn-p2b}
\ee
with $\sigma = {\rm sign}(g)$.
The scaled variable $y$ is negative in the classically allowed region
$Fx < \mu$ such that the wavefunction is oscillatory. In the classically
forbidden region $ Fx > \mu$ one has $y > 0$ and the wavefunction vanishes
as $\phi(y) \sim k \Ai(y)$.
Note that the differential equation (\ref{eqn-p2b}) does not depend
on the nonlinear parameter $g$ explicitly - this dependence is hidden
in the normalization of $\phi_k(y)$. Rescaling the normalization
condition $\| \psi \|^2 = 1$ yields
\be
  2 \int_{y(x=0)}^{+\infty}|\phi_k(y)|^2 \rd y = \frac{|g|}{(2F)^{1/3}} \, .
  \label{eqn-norm-scaled}
\ee

The quantisation condition can now be deduced from the asymptotic form
(\ref{eqn-p2-asymp2}) of the Painlev\'e transcendent.
Note that the definition of a quantum number is not so straightforward as
in the linear case, as new nonlinear eigenstates can emerge and disappear
if the nonlinearity $g$ is changed (see, e.g. \cite{Dago02}).
However, if we restrict ourselves to the nonlinear eigenstates with a
linear counterpart and thus a defined parity, the quantum number can be
identified with the number of zeros of the wavefunction.
Thus the relevant quantisation condition is that the wavefunction
$\psi_n(x)$ must have $n$ zeros. Due to the (anti)symmetry
$\psi(x) = (-1)^n \psi(-x)$, the wavefunction assumes an extremum
($n$ even) or a zero ($n$ odd) at $x=0$.

Using the asymptotic form (\ref{eqn-p2-asymp2}) of the Painlev\'e
transcendent, this condition can now be cast into an explicit form.
As the asymptotic form of the $P_{II}$ transcendent is basically
given by a sine function, a condition for the argument of this sine
function directly follows from the conditions on the wavefunction.
In fact, the argument of the sine at $y(x = 0) =  - 2^{1/3}  \mu / F^{2/3}$
must equal $(n+1)\pi/2$. Inserting this into equation (\ref{eqn-p2-asymp2})
yields the relevant quantisation condition
\be
  \frac{(2 \mu )^{3/2}}{3F} - \dfrac{3}{4} \sigma d^2(k_n)
  \ln\left( \frac{2^{1/3}  \mu}{F^{2/3}}  \right)
   - \theta(k_n) = \frac{n+1}{2} \pi,
   \label{eqn-quant-wedge}
\ee
where $d(k_n)$ and $\theta(k_n)$ are given by equations
(\ref{eqn-p2-connect-const1}) and (\ref{eqn-p2-connect-const2}),
respectively. The advantage of this method is that the problem of solving
a nonlinear boundary value problem is reduced to a single algebraic equation.

However, calculating a nonlinear eigenstate with quantum number $n$
for a given value of the nonlinear parameter $g$ is not so easy.
In fact one has to determine the chemical potential $\mu$ so that
the quantisation condition (\ref{eqn-quant-wedge}) and the normalization
condition (\ref{eqn-norm-scaled}) are fulfilled {\it simultaneously}.
This can be achieved by an iterative method.
It is much easier, however, to start from a {\it fixed} value of $\mu$.
The quantisation condition (\ref{eqn-quant-wedge})
then yields solutions $k_n$ for different quantum numbers $n$. Given
these values of $k_n$, one can calculate the Painlev\'e functions
$\phi_k(y)$ and the effective nonlinear parameter $g_n(\mu)$ from
the normalization integral (\ref{eqn-norm-scaled}).
Rescaling the variables to $x$ and $\psi$ again directly gives the
wavefunction $\psi(x)$.

\begin{figure}[t]
\centering
\includegraphics[width=6cm,  angle=0]{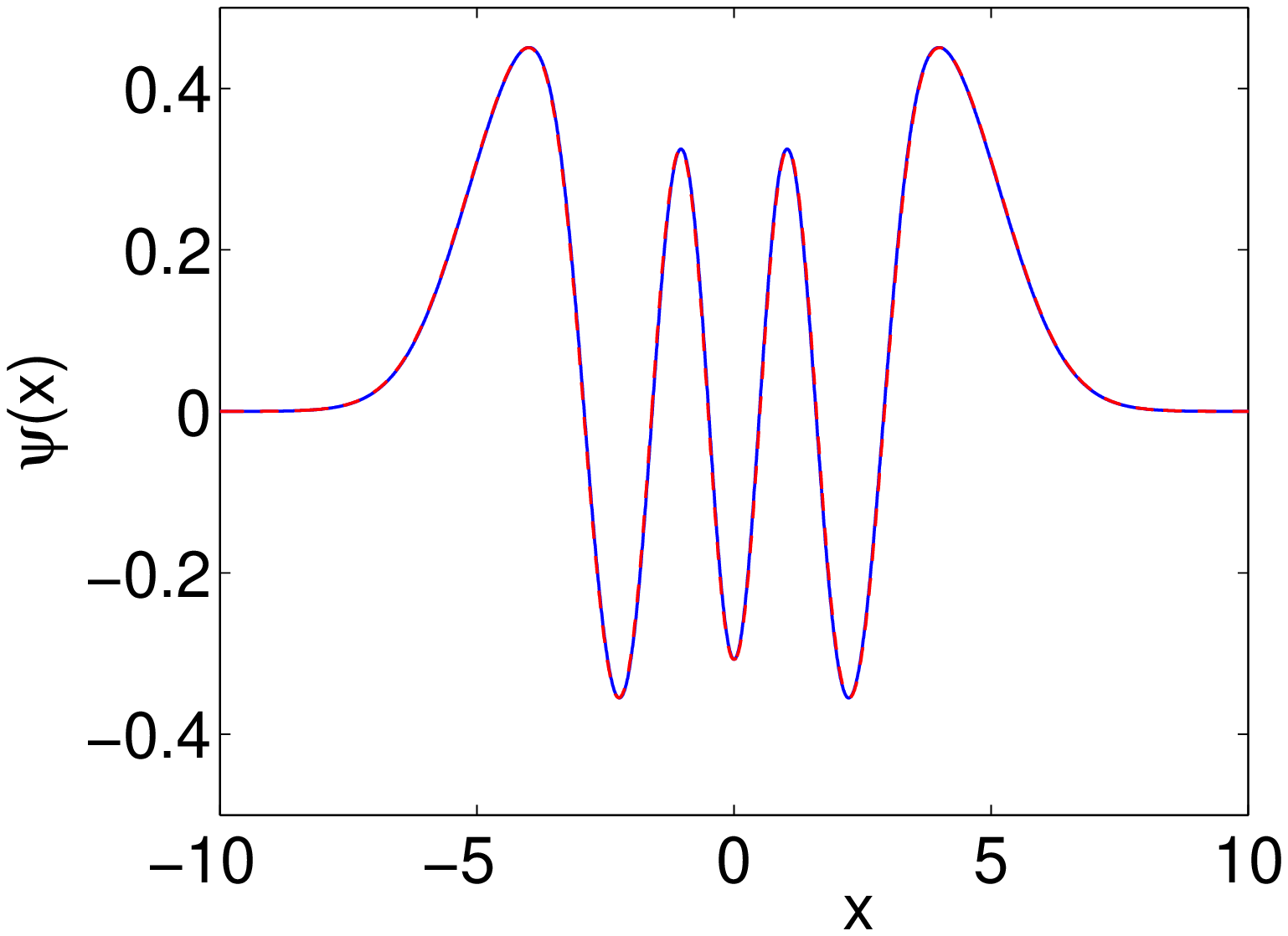}
\hspace{5mm}
\includegraphics[width=6cm,  angle=0]{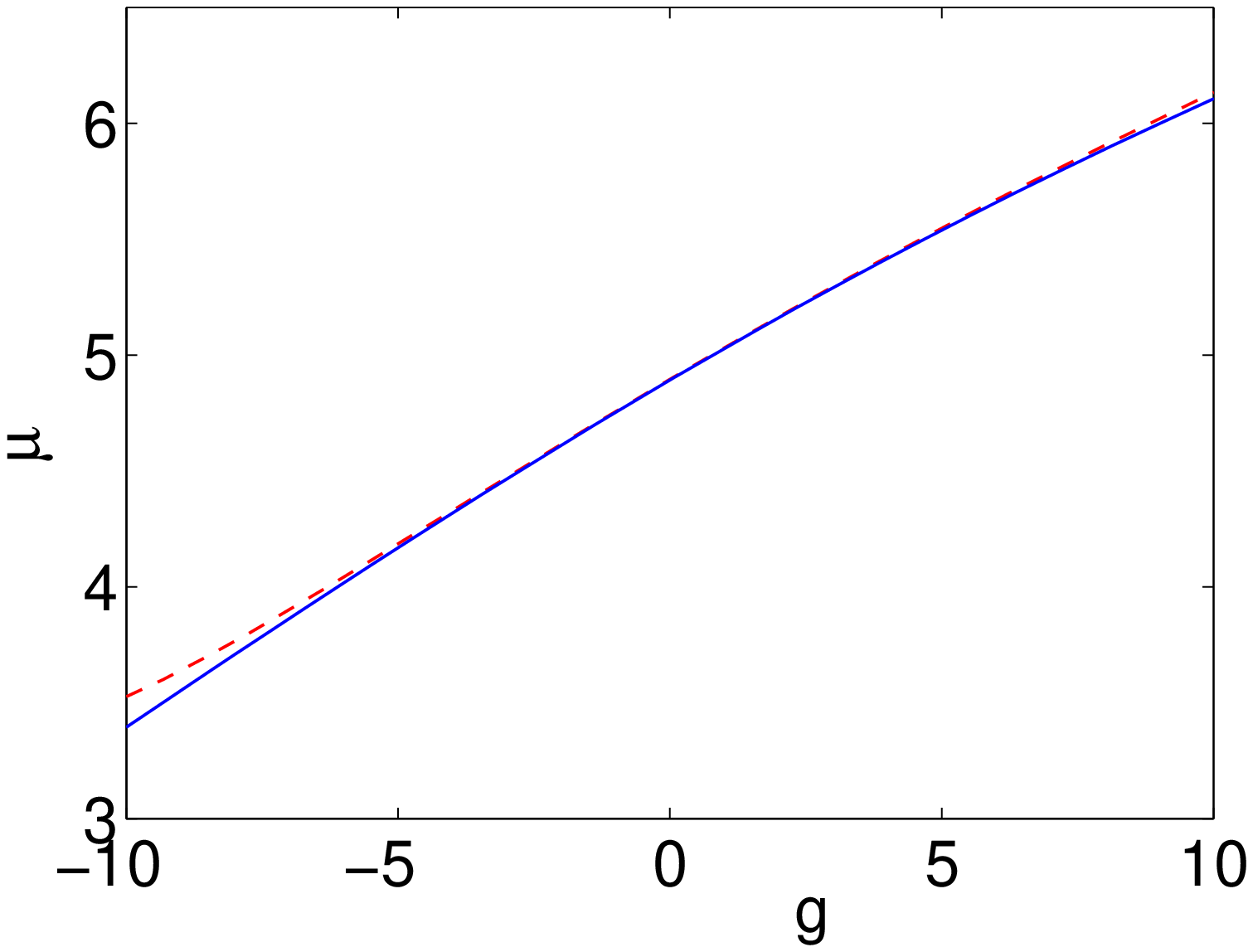}
\caption{\label{fig-wedge_n=6}
Nonlinear eigenstate with quantum number $n = 6$ of the NLSE for a
wedge potential $V(x) = |x|$. Left: wavefunction for $g = 5$, right:
Dependence of the chemical potential on the nonlinearities $g$.
The semiclassical results (dashed red line) are compared to numerically
exact results (solid blue line).}
\end{figure}

To test the feasibility of this approach, we consider the nonlinear
eigenstate $n=4$ for a wedge potential with $F = 1$.
The resulting wavefunction is shown in figure \ref{fig-wedge_n=6}
on the left-hand side (dashed red line) in comparison with the
numerically exact solution (solid blue line). Both wavefunctions
are indistinguishable on the scale of drawing.
The right side shows the dependence of the chemical potential
on the nonlinearity $g$, again in comparison to the numerically
exact values. One observes a good agreement.
The numerical results for the NLSE solutions were obtained using
the standard boundary-value solver {\tt bvp4c} of MATLAB.

\begin{figure}[t]
\centering
\includegraphics[width=7cm,  angle=0]{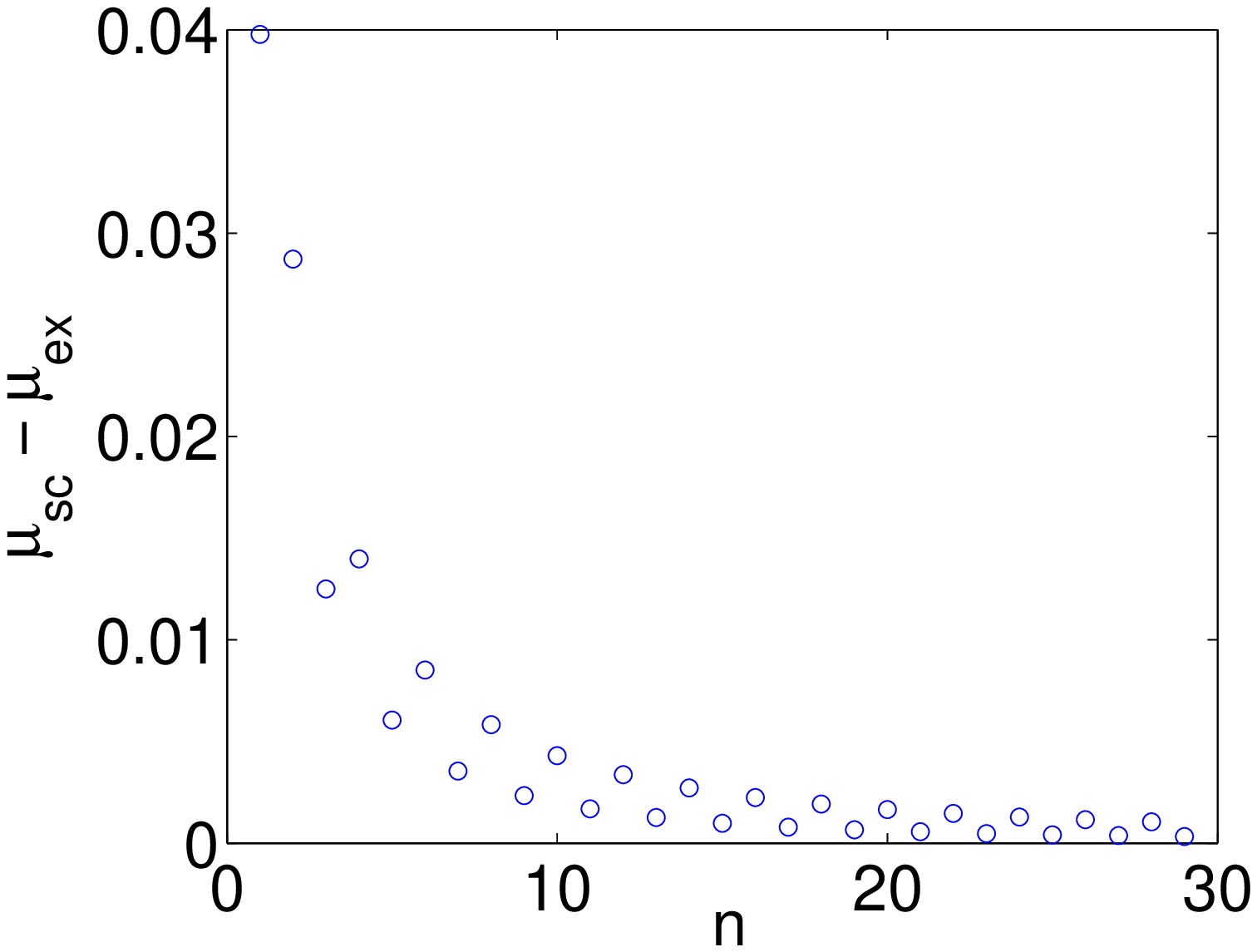}
\caption{\label{fig-wedge_mun}
Solution of the NLSE for a wedge potential: Error of the semiclassical
calculation of the chemical potential $\mu_{sc} - \mu_{ex}$ in
dependence of the quantum number $n$ for $g = 5$.}
\end{figure}

The only error in this calculation results from the replacement of the
$P_{II}$ transcendent by its asymptotic form (\ref{eqn-p2-asymp2}).
This error vanishes rapidly for larger quantum numbers $n$, which
is illustrated in figure \ref{fig-wedge_mun}.
The extrema of the $P_{II}$ transcendent are given less accurately
by the asymptotic form than the zeros. Thus the error is larger for
even quantum number $n$.

\section{The harmonic potential}
\label{sec-harmosc}

Now we want to extend the quantisation method presented in the
previous section to a more important application - the harmonic
trap
\be
  V(x) = \frac{x^2}{2} .
\ee
A common method used in semiclassics is a comparison of the
Schr\"odinger equation to a well-known differential equation,
such as Airy's equation \cite{Mill68,Berr72a}.
Similarly we will map the NLSE for the harmonic trap to the the
$P_{II}$ equation.

We use the mapping ansatz
\be
  \psi(x) = a f(x) \phi(y(x)),
  \label{eqn-ansatz-map}
\ee
well known for the linear Schr\"odinger equation \cite{Berr72a},
however with an additional scaling constant $a$. Differentiating
twice gives
\be
   \frac{1}{a} \frac{\rd^2 \psi}{\rd x^2} = \frac{\rd^2 f}{\rd x^2} \phi
  + 2 \frac{\rd f}{\rd x}  \frac{\rd \phi}{\rd y} \frac{\rd y}{\rd x}
  + f \frac{\rd \phi}{\rd y} \frac{\rd^2 y}{\rd x^2}
  + f \left(  \frac{\rd y}{\rd x} \right)^2 \frac{\rd^2 \phi}{\rd y^2}
  \label{eqn-ansatz-2diff}
\ee
We demand that the terms proportional to $\rd \phi /\rd y$ cancel,
which leads to the condition
\be
  2 \frac{\rd f}{\rd x} \frac{\rd y}{\rd x}  + f \frac{\rd^2 y}{\rd x^2} = 0
  \qquad \Rightarrow  \; f(x) = \left(\frac{\rd y}{\rd x}\right)^{-1/2}.
\ee
Furthermore the term proportional to $\rd^2 f  /\rd x^2$  is assumed to be small
and can be neglected.
Substituting the $P_{II}$ equation (\ref{eqn-p2}) and the NLSE (\ref{eqn-nlse-standard})
into equation (\ref{eqn-ansatz-2diff}) finally yields
\be
  \left(\frac{\rd y}{\rd x}\right)^2 \left[ 2 \sigma \phi^3 + y \phi \right]
  + q^2(x) \phi - 2 g a^2 \left(\frac{\rd y}{\rd x}\right)^{-1} \phi^3 = 0.
\ee
In the linear world, which is given by $g=0$ or $\sigma = 0$ respectively,
this directly gives a differential equation that determines the mapping
$y(x)$.
If the nonlinear effects are small, we can neglect the nonlinear terms
in the mapping equation which yields
\be
  \left(\frac{\rd y}{\rd x}\right)^2 = \frac{-q^2(x)}{y(x)} \, .
  \label{eqn-p2map-dgl}
\ee
The scaling constant $a$ is now chosen such that the error due to the
neglect of the nonlinear terms in the mapping equation (\ref{eqn-p2map-dgl})
is as small as possible.
In the fashion of a least squares fit, $a$ is chosen such that the error
\be
  \chi^2 = \int_{-\infty}^\infty \left[ \sigma \left(\frac{\rd y}{\rd x}\right)^2 - g a^2
  \left(\frac{\rd y}{\rd x}\right)^{-1} \right]^2 \phi(y(x))^6 \rd x
  \label{eqn-map-nonlin-error}
\ee
is minimal. This can be done at the end of the calculation, after $\phi(y(x))$
has been determined.

\begin{figure}[t]
\centering
\includegraphics[width=6cm,  angle=0]{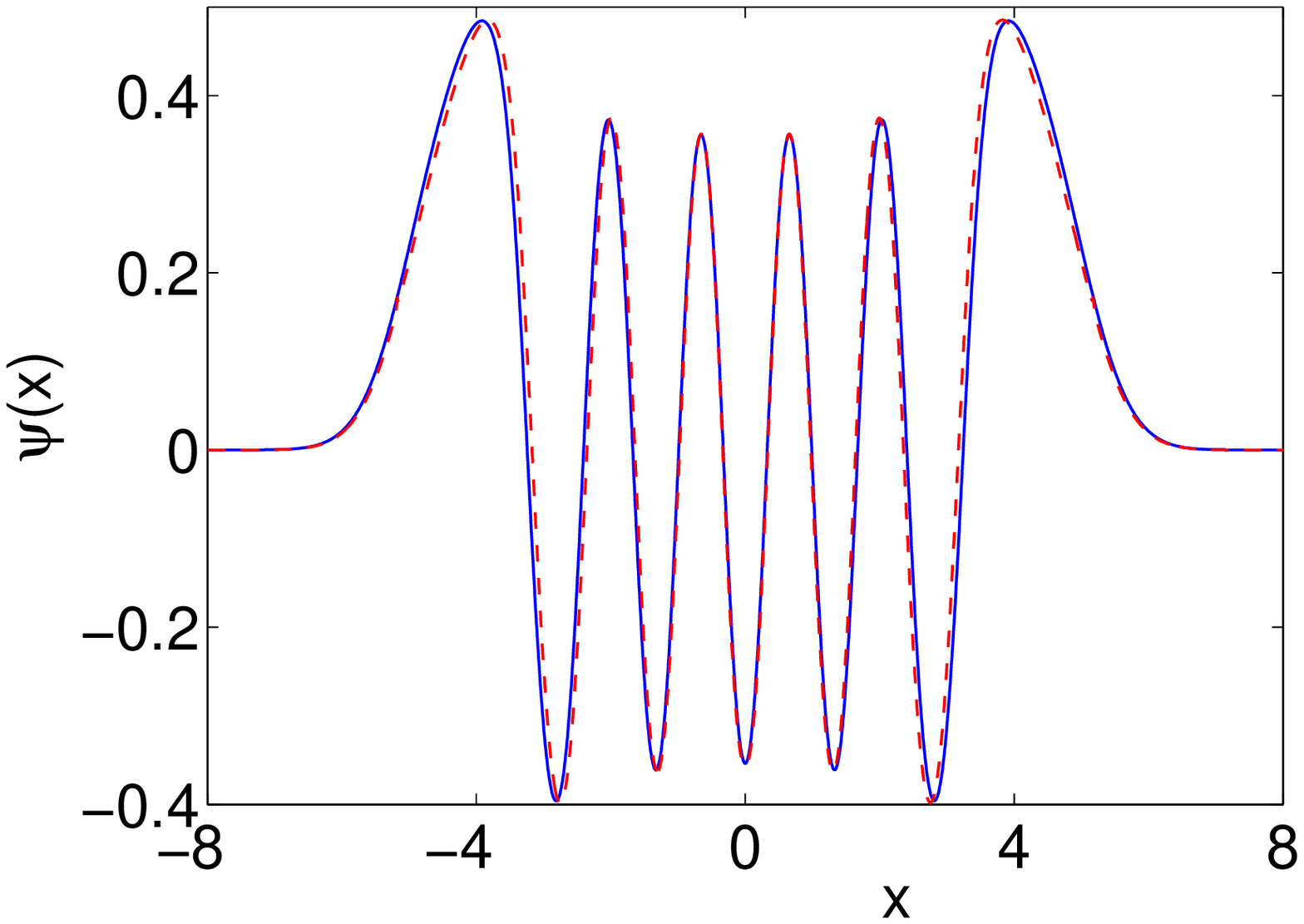}
\hspace{5mm}
\includegraphics[width=6cm,  angle=0]{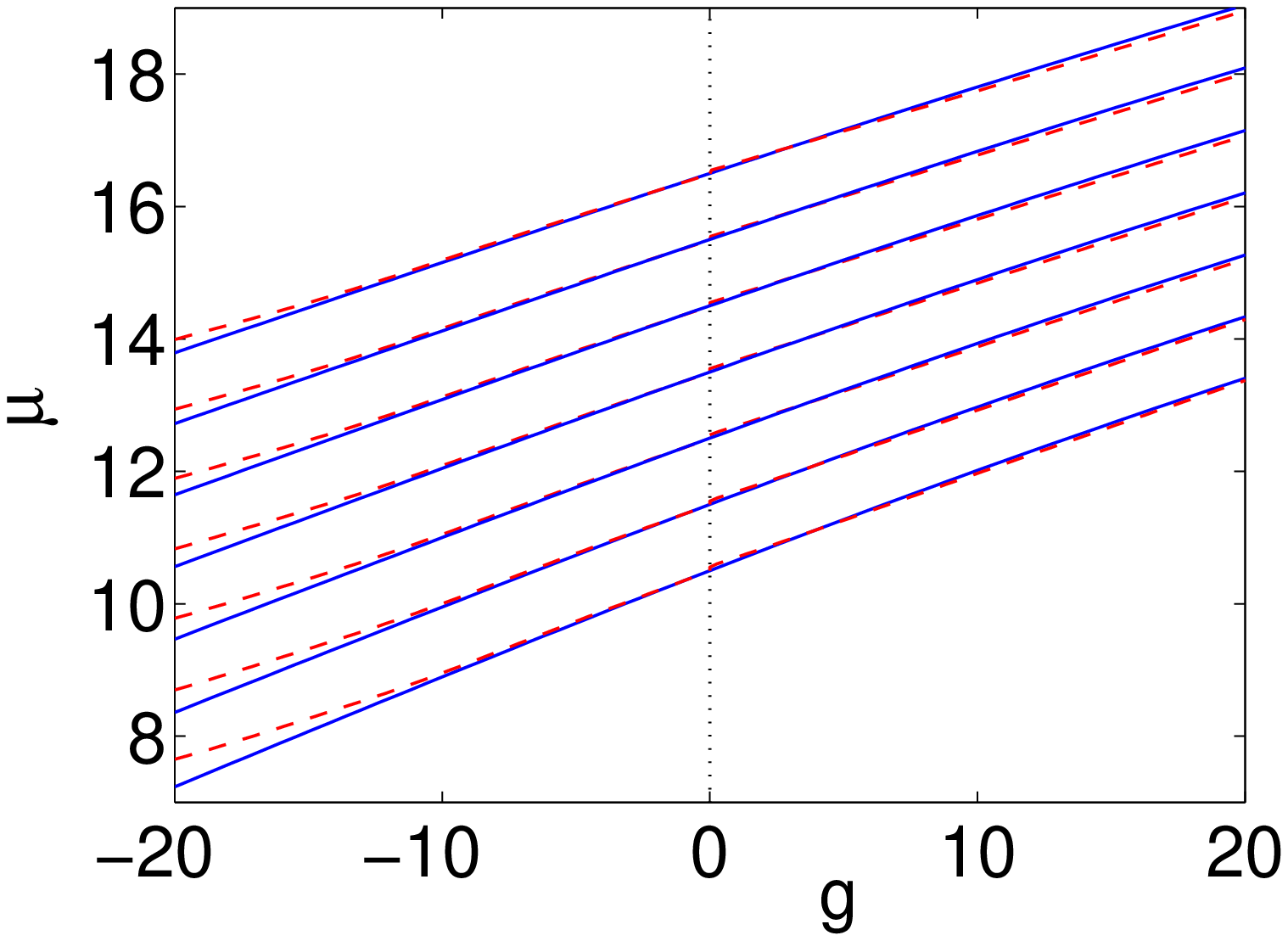}
\caption{\label{fig-p2map_ho_n=10}
Nonlinear eigenstates of the NLSE in a harmonic potential $V(x) = x^2/2$.
Left: Wavefunction of the eigenstate with quantum number $n=10$ for $g = 10$,
right: Dependence of the chemical potential on the nonlinearities $g$
for the eigenstates $n =10-16$.
The semiclassical results (dashed red line) are compared to numerically
exact results (solid blue line).}
\end{figure}

This mapping can now be used to approximately calculate eigenstates in symmetric
single minimum potentials at $x = 0$, e.g. a harmonic trap $V(x) = x^2/2$.
For wavefunctions with a linear counterpart, that have a defined parity,
we can restrict our analysis to $x \ge 0$.
To avoid a divergence at the classical turning point $x_t$ the mapping has to be
such that $y(x_t) = 0$. Thus the integrating equation (\ref{eqn-p2map-dgl})
yields the mapping in explicit form
\be
  y(x)  = \pm \left[ \pm \frac{3}{2} \int_{x_t}^x \sqrt{|q^2(x')|} \rd x' \right]^{2/3},
  \label{eqn-ho-map-exp}
\ee
where the $-$ sign is taken in the classically allowed region $x < x_t$ and the
$+$ sign is taken in the classically forbidden region $x > x_t$.

The quantisation condition is deduced from the asymptotic form of the Painlev\'e
transcendent (\ref{eqn-p2-asymp2}) exactly as in section \ref{sec-wedge}.
The only difference is that the mapping is now given by equation
(\ref{eqn-ho-map-exp}), such that the expression for $y(x = 0)$ is a
bit more complicated.
Thus the relevant quantisation condition is given by
\be
  \dfrac{2}{3} |y(0)|^{3/2} - \dfrac{3}{4} \sigma d^2(k_n) \ln |y(0)|
   - \theta(k_n) = \frac{n+1}{2} \pi.
\ee

To test the feasibility of this approach, we consider the nonlinear
eigenstates for a harmonic potential $V(x) = x^2/2$. The result for the
eigenfunctions with $n = 10$ are shown in figure \ref{fig-p2map_ho_n=10}.
The left-hand side shows the wave function calculated using the mapping procedure
(dashed red line) in comparison with the numerically exact solution (solid
blue line). The right-hand side shows the dependence of the chemical potential
on the nonlinearity $g$, again in comparison to the numerically
exact values. One observes a good agreement.

\begin{figure}[t]
\centering
\includegraphics[width=7cm,  angle=0]{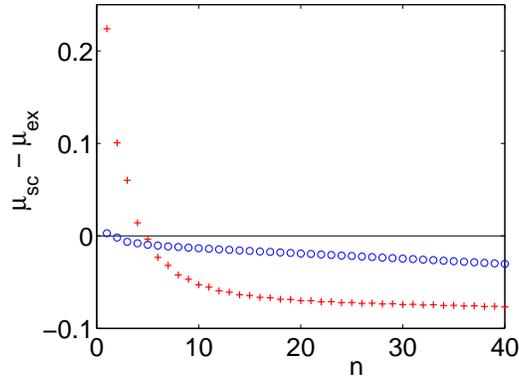}
\caption{\label{fig-p2map_ho_mun}
Solution of the NLSE for a harmonic potential: Error of the semiclassical
calculation of the chemical potential $\mu_{sc} - \mu_{ex}$ in
dependence of the quantum number $n$ for $g = 1$ (blue circles)
and $g = 10$ (red crosses).}
\end{figure}

Figure \ref{fig-p2map_ho_mun} shows results for different quantum numbers $n$.
The error of the semiclassical calculation, i.e. the difference of the semiclassical
value for the chemical potential $\mu_{sc}$ and the numerically exact value $\mu_{ex}$
is plotted against the quantum number $n$ for $g = 1$ and $g =  10$.
Except for very small values of $n$ and $g = 10$, for which the reduction to
the asymptotic form (\ref{eqn-p2-asymp2}) is not valid, one
obtains reasonable results for the semiclassical approximation.

The method introduced above can be extended to asymmetric trapping
potentials. Then one has to construct solutions around the two
classical turning points separately, which are matched at a
'mid-phase point' \cite{Mill68}.
In this spirit the restriction to symmetric or anti-symmetric solutions
above is nothing but a matching of two solutions at the mid-phase point
$x=0$.

In the linear case $g=0$ a mapping to a similar potential with
two classical turning points, in fact the harmonic potential,
avoids this matching procedure \cite{Berr72a}.
In the nonlinear case, however, the single-turning point equation
$P_{II}$ has some advantages compared to the NLSE with a harmonic
potential because the $P_{II}$ equation is free of movable branch
points and connection formulae are well known.

\section{Mapping to a constant potential}
\label{sec-map-const}

It is well-known that the free NLSE 
\be
   \frac{\rd^2 \phi}{\rd y^2}  + 2 \nu \phi(y) - 2 g \phi(y)^3 = 0
   \label{eqn-free_nlse}
\ee
has soliton solutions. Bright solitons are found for $\nu < 0$ and $g < 0$,
given by
\be
  \phi(y) = \sqrt{2\nu / g} \, \sech\left( \sqrt{-2\nu} (y - y_0) \right),
   \label{eqn-bsol-free}
\ee
and dark solitons for $\nu < 0$ and $g > 0$ are given by
\be
  \phi(y) = \sqrt{\nu / g} \tanh\left( \sqrt{\nu} (y - y_0) \right).
\ee

\begin{figure}[t]
\centering
\includegraphics[width=7cm,  angle=0]{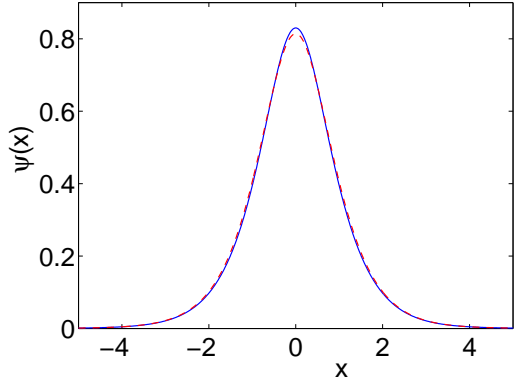}
\caption{\label{fig-c2map-bsol-wf}
Bright soliton solution in a cosine-potential of strength $w = -0.2$ for
$\mu = -1$. Numerically exact results (solid blue line) are compared to
results from the mapping technique (dashed red line).}
\end{figure}

Using a mapping technique as in the previous section we explore the
effects of a small additional cosine potential on these solitons.
In fact we consider the NLSE
\be
   \frac{\rd^2 \psi}{\rd x^2}  + 2 (\mu - w \cos(x)) \psi(x) - 2 g \psi(x)^3 = 0
\ee
Using again the ansatz (\ref{eqn-ansatz-map}) and following the lines of reasoning
of section \ref{sec-harmosc}, one arrives at
\be
  \left(\frac{\rd y}{\rd x}\right)^2 \left[ 2 g \phi^3 - 2 \nu \phi \right]
  + 2(\mu - v \cos(x) ) \phi - 2 g a^2 \left(\frac{\rd y}{\rd x}\right)^{-1} \phi^3 = 0
  \label{eqn-map-all-const}
\ee
Again one chooses the scaling factor $a$ to minimize the difference of the nonlinear
terms
\be
  \chi^2 = \int_{-\infty}^\infty \left[ g \left(\frac{\rd y}{\rd x}\right)^2 - g a^2
  \left(\frac{\rd y}{\rd x}\right)^{-1} \right]^2 \phi(y(x))^6 \, \rd x
\ee
and neglects them in equation (\ref{eqn-map-all-const})
to arrive at the mapping equation
\be
  \left(\frac{\rd y}{\rd x}\right)^2 = \frac{\mu - w \cos(x)}{\nu}
\ee
To ensure that the right-hand side is positive, one must always be in the
classically allowed region ($\mu > w$, thus $\nu > 0$) or the classically
forbidden region ($\mu < - w$, thus $\nu < 0$); values in the interval
$\mu \in (-w,w)$ cannot be treated within this framework.

To show the validity of this method we calculate a bright soliton solution
in a cosine lattice $V(x) = w \cos(x)$. Figure \ref{fig-c2map-bsol-wf}
shows the wavefunction calculated by the mapping method in comparison to
the numerically exact solution. One observes a good agreement.

\begin{figure}[t]
\centering
\includegraphics[width=6cm,  angle=0]{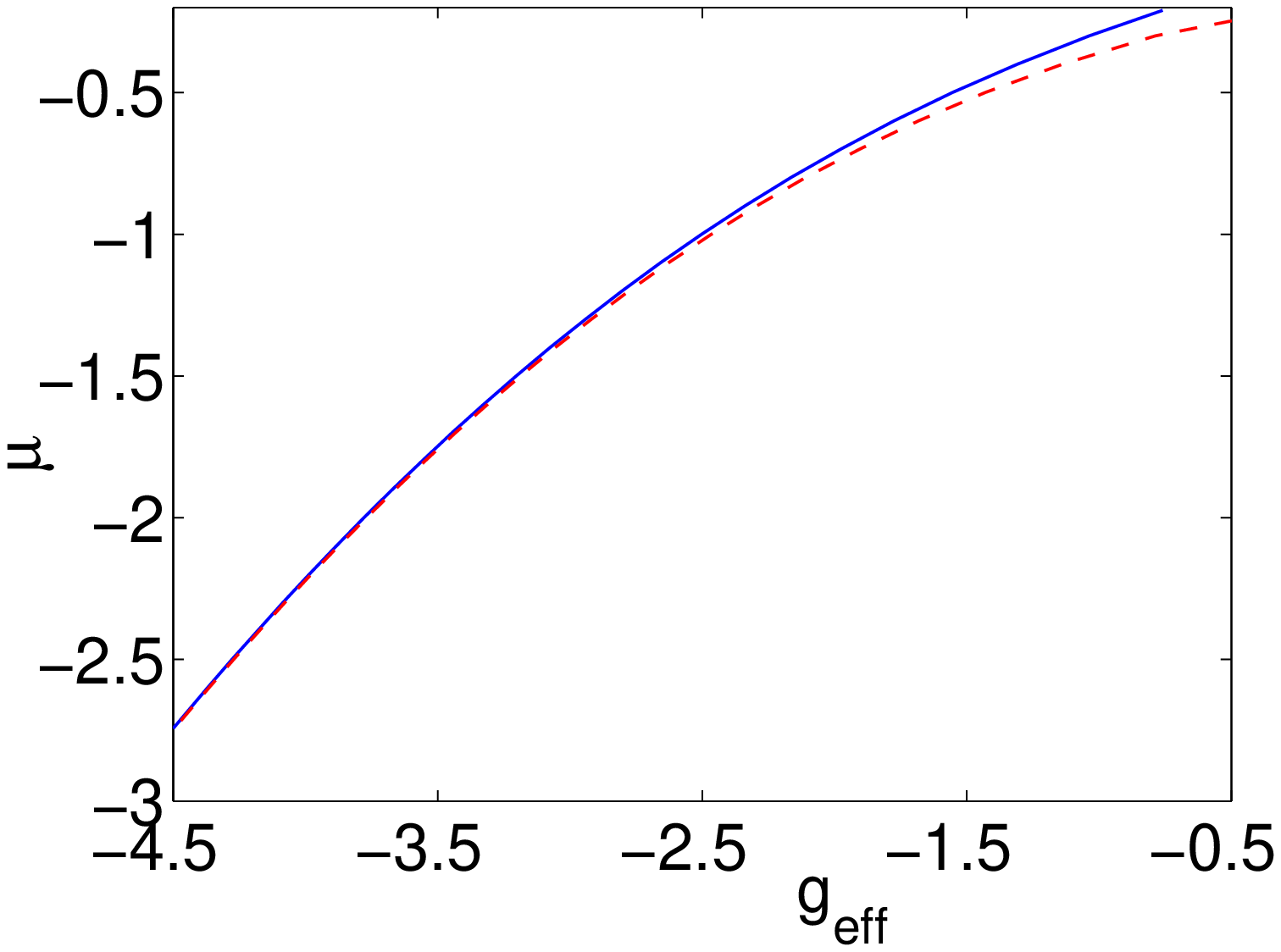}
\hspace{5mm}
\includegraphics[width=6cm,  angle=0]{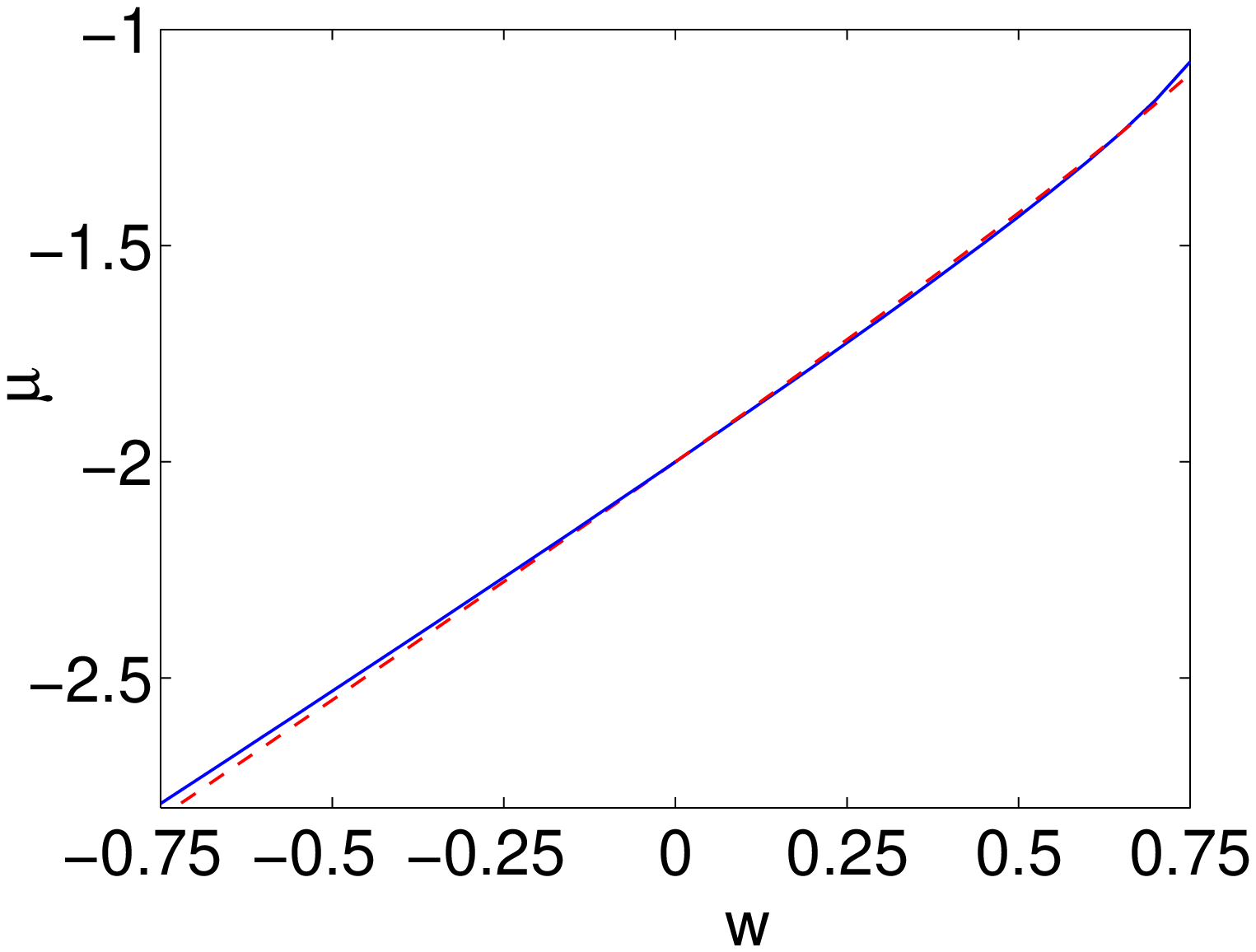}
\caption{\label{fig-c2map-bsol-mu}
Shift of the chemical potential of a bright soliton in a cosine-potential
in dependence of the nonlinearity $g_{\rm eff}$ for $w = -0.2$ (left)
and in dependence of the potential strength $w$ for $g_{\rm eff} - 4$
(right). Numerically exact results (solid blue line) are compared to
results from the mapping technique (dashed red line).}
\end{figure}

Furthermore we calculate the dependence of the chemical potential potential $\mu$
of such a bright soliton on the nonlinarity $g_{\rm eff}$ for a fixed value of $w$
and on the potential strength $w$ for a fixed nonlinearity $g_{\rm eff}$.
The results obtained by the mapping method and the numerically exact results
are compared in figure \ref{fig-c2map-bsol-mu}.
One observes a good agreement.


\section*{Acknowledgements}
Support from the Studienstiftung des deutschen Volkes
and the Deutsche Forschungsgemeinschaft via the Graduiertenkolleg
``Nichtlineare Optik und Ultrakurzzeitphysik'' is gratefully acknowledged.
We thank R. S. Kaushal for stimulating discussions.


\end{document}